\documentclass[12pt,a4paper]{article}
\usepackage{amssymb}

\newcommand{\be}{\begin{equation}}
\newcommand{\ee}{\end{equation}}
\newcommand{\ba}{\begin{eqnarray}}
\newcommand{\ea}{\end{eqnarray}}

\newcommand{\NP}[1]{Nucl.\ Phys.\ {\bf #1}}
\newcommand{\PL}[1]{Phys.\ Lett.\ {\bf #1}}

\newcommand{\nonum}{\nonumber \\[1.5mm]}

\newcommand{\R}{\mbox{\rm I\hspace{-.4ex}R}}

\newcommand{\lb}{\lambda}
\newcommand{\eps}{\epsilon}
\newcommand{\dd}{{\partial}}
\newcommand{\ra}{{\rightarrow}}
\newcommand{\rra}{{\longrightarrow}}
\newcommand{\cH}{{\cal H}}

\newcommand{\nl}{\mbox{[\hspace{-.35ex}[}}
\newcommand{\nr}{\mbox{]\hspace{-.32ex}]}}

\newcommand{\fbar}{{\overline{f}}}

\newcommand{\hbbar}{{\overline{h}}}

\begin{document}

\begin{titlepage}
\renewcommand{\thefootnote}{\fnsymbol{footnote}}
\mbox{}
\vspace{1.5cm}

\begin{center}
{\Large \bf A fixed point for truncated Quantum}\\[3mm]
{\Large \bf Einstein Gravity}
\vspace{2.5cm}

{\large P.~Forg\'acs and M. Niedermaier%
\footnote{e-mail: {\tt max@phys.univ-tours.fr}}}
\\[8mm]
{\small\sl Laboratoire de Mathematiques et Physique Theorique}\\
{\small\sl CNRS/UMR 6083, Universit\'{e} de Tours}\\
{\small\sl Parc de Grandmont, 37200 Tours, France}\\
\vspace{3cm}

{\bf Abstract}
\end{center}
\vspace{-5mm}
\begin{quote}
A perturbative quantum theory of the two Killing vector reduction of 
Einstein gravity is constructed. Although the reduced theory inherits 
from the full one the lack of standard perturbative renormalizability, 
we show that strict cutoff independence can be regained to all loop orders
in a space of Lagrangians differing only by a field dependent conformal 
factor. A closed formula is obtained for the beta functional governing 
the flow of this conformal factor. The flow possesses a unique fixed 
point at which the trace anomaly is shown to vanish. The approach 
to the fixed point is compatible with Weinberg's ``asymptotic safety'' 
scenario. 
\end{quote}
\vfill

\setcounter{footnote}{0}
\end{titlepage}


The non-renormalizability of the Einstein-Hilbert action in a 
perturbative expansion around a fixed background spacetime 
\cite{Weinberg,tHVelt74,GorSagn86} is often taken to indicate that 
a quantum theory based on it is meaningful only as an effective field theory 
with a domain of applicability that is limited both in the infrared 
and in the ultraviolet (UV) regime. Quantitatively however the 
corrections in the effective field theory realm are way too small 
to be physically relevant \cite{Dono94}. ``Large'' quantum gravity  
effects, if any, require a much higher degree of universality 
viz insensitivity on the cut-off and computational procedures. 
Early on S.~Weinberg \cite{Weinberg} stressed that in a Wilsonian 
framework this would (hypothetically) be the case if the averaging 
flow had a hidden UV stable fixed point, whose critical manifold 
might still be finite dimensional. This would be equivalent to 
quantum Einstein gravity being non-perturbatively renormalizable! 
-- a scenario whose verification or refutation is clearly of outstanding 
theoretical significance. Some version of this ``asymptotic safety'' 
scenario arguably underlies also the dynamical triangulations approach 
to quantum gravity and, at least indirectly, the canonical quantization   
program of A.~Ashtekar and its ramifications. The scenario lay 
mostly dormant since \cite{Weinberg}, presumably for lack of 
evidence and computational techniques. Recently however important 
evidence for its viability has been reported by M.~Reuter and 
O.~Lauscher \cite{Reuter,ReuterLausch}. Their setting is that of 
`exact' flow equations for an averaged effective action functional. 
Due to the complexity of the problem only the (extended) 
Einstein-Hilbert truncation has been studied, corresponding to a 
projection of the averaging flow into a 2 (or 3) dimensional subspace 
in which a non-Gaussian fixed point is seen \cite{ReuterLausch}.

Here we report on fairly complete results obtained on analogous issues
in a truncation of Einstein gravity describing spacetimes with two 
Killing vectors. The reduced system turns out to inherit the 
non-renormalizability of full Einstein gravity (beyond one loop).  
The renormalization flow will therefore move in an {\it infinite} 
dimensional space of Lagrangians constrained only by the initial symmetry 
requirement. In this space we are able to regain UV renormalizability 
in the sense of strict cutoff independence.  
The space turns out to be parameterized by functions $h(\,\cdot\,)$ of
one real variable which generalize the notion of a renormalized coupling.
Functions connected by the renormalization flow $\mu\, \ra \,
\hbbar(\,\cdot\,,\mu)$ defined below are regarded as equivalent. 
This infinite dimensional space is now the appropriate arena 
to search for a fixed point (function) and, remarkably, there is one! 
Moreover it is unique and the renormalization flow appears to be 
driven locally toward the fixed point. Of course one cannot expect that 
``truncation of classical degrees of freedom'' and ``quantization'' 
are commuting operations. Based on the `anti-screening' nature of 
the interaction \cite{Dono94} it is however hard to imagine that a 
symmetry reduction of classical degrees of freedom would make 
the UV renormalization properties worse. In other words, 
our analysis also provides a complementary test of the 
asymptotic safety scenario: If quantum Einstein gravity indeed has 
a hidden fixed point it should emerge already in the symmetry 
reduced quantum theory (but not vice versa).

We begin with describing the outcome of the classical reduction 
procedure \cite{MatzMis67,Ernst}. One is interested in the solutions of 
Einstein's equations with two commuting Killing vectors. 
They cover a variety of physically interesting situations: If one of the 
Killing vectors is timelike these are stationary axisymmetric 
spacetimes, among them in particular all the prominent black hole 
solutions. If both Killing vectors are spacelike the subsector comprises 
(depending on a certain signature) cylindrical gravitational
waves, colliding plane gravitational waves, as well as generalized Gowdy 
cosmologies. The set of all these solution is equipped with a 
symplectic structure which is induced by the standard symplectic
structure of general relativity. It is important that both the 
reduced Einstein equations and the induced symplectic structure
follow from a two-dimensional (2D) action principle, which we 
take as the basis for developing the quantum theory.  This action 
couples 2D gravity non-minimally 
via a ``radion'' field $\rho$ to a 2D matter system. The radion 
field is related to the determinant of the internal metric, 
the matter system is a noncompact ${\rm O}(1,2)$ nonlinear sigma-model. 
This means the fields are maps $n = (n^0,n^1,n^2)$ from the 
2D spacetime manifold into the hyperboloid $H_{\eps} = \{n = (n^0,n^1,n^2) 
\in  \R^{1,2}\,|\, n\cdot n = (n^0)^2 - (n^1)^2 - (n^2)^2 = \eps\}$.
The sign $\eps = \pm 1$ distinguishes the two main situations,
where either both Killing vectors are spacelike ($\eps =+1$) or 
only one of them ($\eps =-1$). Accordingly the 2D metric $\gamma_{\mu\nu}$ 
will either have Lorentzian or Euclidean signature; without 
(much) loss of generality we shall always assume it to be 
conformally flat $\gamma_{\mu\nu} \sim e^{\sigma} \eta_{\mu\nu}$,
and describe the dynamics in terms of $\sigma$. The phase space 
is then characterized by the equations of motion and the symplectic 
structure following from the flat space action 
$S = \int \!d^2 x L$, with 
\be
L(n, \rho, \sigma) = -\frac{1}{2 \lb}[ \rho 
\dd^{\mu} n \cdot \dd_{\nu} n + \eps \dd^{\mu} \rho \dd_{\mu}
(2 \sigma + \ln \rho)]\;,
\label{i1} 
\ee
where $n\cdot n = \eps$ and $\lb$ is Newton's constant per unit 
volume of the internal space. Further, to account for the original 
2D diffeomorphism invariance, the weak vanishing of the 
hamiltonian and the 
diffeomorphism constraints $\cH_0 \approx 0,\,\cH_1 \approx 0$
has to be imposed. The latter are given by $\cH_0 = T_{00}$ and $\cH_1 = 
T_{01}$, if $T_{\mu\nu}$ denotes the classical energy momentum 
tensor derived from (\ref{i1}). The trace $T^{\mu}_{\;\mu}$ 
vanishes on-shell. Classical observables are quantities 
which Poisson commute with these constraints and are weakly 
nonvanishing. A unique feature of the 2-Killing reduction is 
that an infinite set of such (nonlocal) observables can be 
constructed explicitly! (They are analogues of the nonlocal 
conserved charges in conventional sigma-models \cite{Lusch}.) 
This makes the system a compelling laboratory for studying quantum 
aspects of general relativity \cite{Asht96,ERwaves,NicKor96,qernst}.

We now want to embark on a Dirac quantization of the 
system. That is the vector $n$ as well as $\rho, \sigma$ are promoted 
to independent quantum fields whose dynamics is governed by the 
action (\ref{i1}). The constraints are intended to be defined 
subsequently as composite operators in terms of which the projection onto 
the physical state space is formulated. Clearly the key issue to be 
addressed is that of the renormalizability of an action functional 
motivated by (\ref{i1}) and its symmetries and extended by 
suitable sources needed to define composite operators.     
In contrast to nonlinear sigma-models without coupling to 
gravity we find that the quantum field theory based on (\ref{i1}) 
is {\it not} UV renormalizable in the strict sense.  
Renormalizability in the strict quantum field theoretical sense  
typically presupposes that the bare and the renormalized 
(source extended) action have the same functional form, only the 
arguments of that functional (fields, sources, and coupling constants) 
get renormalized. Though the bare action is motivated by 
the classical one it can be very different from it. In any case the 
form of the (bare=renormalized) action functional is meant to be known 
before one initiates the renormalization. The 2-Killing 
vector reduction of general relativity has been known for
some time to be renormalizable at the 1-loop level \cite{dWGNR92}.
Here we confirm this result, but we also find that the 
system is not renormalizable in the above sense beyond  one loop.
This means the reduced system accurately portrays the features of 
full Einstein gravity \cite{tHVelt74,GorSagn86}. Technically the 
non-renormalizability arises because the dependence 
on the dimensionless radion field $\rho$ is not constrained 
by any Noether symmetry. Thus it can -- and does -- enter 
the counter terms in a different way as in the bare action,    
no matter how the latter is chosen.

The solution we propose is to renormalize the theory in a space of 
Lagrangians which differ by an overall conformal factor that is a function 
of $\rho$. More precisely we showed that to {\it all orders} in 
the loop expansion nonlinear field renormalizations exist such 
that for any prescribed bare $h_B(\,\cdot\,)$ there exists
a renormalized $h(\,\cdot\,)$ such that       
\ba
\frac{h_B(\rho_B)}{\rho_B} L(n_B,\rho_B,\sigma_B) &=&  
\frac{h(\rho)}{\rho} L(n,\rho,\sigma)\;,
\nonum 
\quad \mbox{but} \quad h_B(\,\cdot\,) &\neq&  h(\,\cdot\,) \,.
\label{i2}
\ea
A subscript `${}_B$' denotes the bare fields while the plain 
symbols refer to the renormalized ones. The fact that 
$h_B(\,\cdot\,)$ and $h(\,\cdot\,)$ differ marks the deviation from 
conventional renormalizability; it holds with one notable exception
described below. In order to be able to perform explicit computations 
we derive this result in a specific computational scheme: 
Dimensional regularization, minimal subtraction and the
covariant background field expansion, with adaptations from  
\cite{Frie85,Osb87}. Ultimately the reason 
why (\ref{i2}) works is that $L$ possesses two conformal symmetries 
in {\it field space} (possibly linked to those in \cite{NicJul}) 
which together with the obvious strict symmetries 
characterize the Lagrangian up to a $\rho$-dependent prefactor.
Denoting by $L_h$ the right hand side of (\ref{i2}) there exist
currents $C_{\mu}$ and $D_{\mu}$ obeying the on-shell identities
\be
\dd^{\mu} C_{\mu} = \rho \dd_{\rho} \ln h \cdot L_h\;,
\quad \quad 
\dd^{\mu} D_{\mu} = -\ln \rho \cdot \rho \dd_{\rho} \ln h 
\cdot L_h\;.
\label{cward}
\ee
They can be promoted to Ward identities for the corresponding 
composite operators and could also serve to characterize the 
theory in a regularization independent way. (See \cite{BBBC88} 
for an analysis of this type for conventional sigma-models). 
We made no attempt to address the infrared problem 
here because, guided by the analogy to the abelian sector \cite{ERwaves},
we expect it to disappear upon projection onto the physical 
state space. Even in the above explicit scheme the details of the 
renormalization are fairly technical and will be 
described in \cite{PTpaper}. Here we only mention three important 
points concerning composite operators, again valid to all orders
of the loop expansion: (i) Arbitrary functions
of $\rho$ are finite as composite operators; no additional 
renormalizations are needed. (ii) The ${\rm O}(1,2)$ Noether 
currents $J_{\mu}^i, \, i=0,1,2$, can be defined as finite 
composite operators and obey $\nl J^i_{\mu}(h,n)\nr = 
J^i_{\mu}(h_B,n_B)$, where $\nl \,\cdot\,\nr$ denotes the 
normal product in the above scheme. (iii) The energy momentum 
operator can be defined as a composite operator $\nl T_{\mu\nu} \nr$;
it is unique up to an improvement term of the form 
$\Delta T_{\mu \nu} = (\dd_{\mu} \dd_{\nu} - \eta_{\mu\nu} \dd^2) 
\Phi$, where the bare and the renormalized improvement potential 
are of the form $\Phi_B = f_B(\rho_B) + f_0 \sigma_B$ and $\Phi = f(\rho) + 
f_0 \sigma$, respectively. Here $f_0$ is a finite constant 
and in general $f_B(\,\cdot\,) \neq f(\,\cdot\,)$.

The function $h(\,\cdot\,)$ plays the role of a generalized 
(renormalized) coupling. In contrast to Newton's constant $\lb$, 
which is an `inessential' coupling in Weinberg's terminology 
\cite{Weinberg} (both in our truncated and in full quantum 
Einstein gravity) the function $h(\,\cdot\,)$ can be seen from 
(\ref{cward}) to be an essential coupling (or rather an infinite 
collection thereof), while $\lb$ does not require renormalization.    
To lowest order it is fixed by strict renormalizability to be 
$h(\rho) = \rho^p + O(\lb)$, for some $p\!\neq\!0$. To this order 
it has a direct physical interpretation. Namely $h(\rho)/\rho$ is 
the conformal factor in a Weyl transformation $G_{MN}(x) \,\ra\, 
\Omega(\rho(x))\, G_{MN}(x)$ of a generic 4D metric $G_{MN}(x)$ with 
two Killing vectors in adapted coordinates. The spacetime dependence 
of $\Omega$ enters only through the scalar field $\rho$, a 
concept familiar from scalar-tensor theories of gravity. Beyond one loop 
one is forced to allow for nontrivial finite deformations $h(\rho) = 
\rho^p + \frac{\lb}{2\pi} h_1(\rho) + \big( \frac{\lb}{2\pi} \big)^2
h_2(\rho) + \ldots$, where $h_l(\rho),\,l \geq 1,$ can be arbitrary functions 
of $\rho$, and for definiteness we take $p\!>\!0$ from now on. The 
deformation is necessary because $h(\,\cdot\,)$ is subject to a 
nontrivial flow equation $\mu \frac{d}{d\mu} \hbbar = \beta_h(\hbbar/\lb)$,
where $\mu$ is the renormalization scale and $\mu \,\ra \,
\hbbar(\,\cdot\,,\mu)$ is the `running' coupling function. 
To avoid a possible misunderstanding let us 
stress that the way how $\hbbar(\,\cdot\,,\mu)$ depends on its 
first argument changes with $\mu$, {\it not} just the value $h(\rho(x))$, 
which is still a function of spacetime and hence not a good coupling.   
The specific form of the corresponding functional beta function 
$\beta_h(h)$ does not allow one to preserve $h(\rho) \sim 
\rho^p$, even if this was the inital choice for some normalization
scale $\mu_0$. However one can impose $h(\rho)/\rho^p \,\ra \,1$ 
for $\rho \,\ra \,\infty$ as a boundary condition; it is preserved 
under the flow and guarantees that the flow is solely driven 
by the counter terms, as it should.

Remarkably $\beta_h(h)$ can be obtained in closed form and is given by 
\be
\beta_{h}(h/\lb) = - \rho \dd_{\rho} 
\left[ \frac{h(\rho)}{\lb} \int_{\rho}^{\infty} 
\frac{du}{u} \frac{h(u)}{\lb} \beta_{\lb} 
\Big(\frac{\lb}{h(u)} \Big) \right]\;. 
\label{i3}
\ee
Here $\beta_{\lb}(\lb)$ is the conventional (numerical) beta 
function of the ${\rm O}(1,2)$ nonlinear sigma-model without
coupling to gravity, computed in the minimal subtraction scheme. 
$\beta_{h}(h)$ can thus be regarded as a ``gravitationally
dressed'' version of $\beta_{\lb}(\lb)$, akin to the phenomenon
found in \cite{KlKoPo93} in a different context.  With $\beta_h(h)$ 
known explicitly we can search for a fixed point function
$h^{\rm beta}(\,\cdot\,)$, satisfying $\beta_h(h^{\rm beta}) =0$.
Most importantly it exists. Further subject to the above 
asymptotic boundary condition it is unique and comes out as 
\be
h^{\rm beta}(\rho) = \rho^p - \frac{\lb}{2\pi} \frac{2 \zeta_2}{\zeta_1} 
- \Big(\frac{\lb}{2\pi}\Big)^2 \frac{3 \zeta_3}{2 \zeta_1} \rho^{-p} + 
\ldots\,,
\label{i4} 
\ee 
where $l \zeta_l,\,l\! \geq \!1$, are the ${\rm O}(1,2)$ beta function 
coefficients: $\zeta_1 = - \eps,\,\zeta_2 =1/2,\,\zeta_3 = \eps 5/12$
\cite{Hikami,Wegner}. Note that both the functional form 
and the numerical coefficients in $h^{\rm beta}(\rho)$ are fixed 
by the ``conformal renormalizability'' condition (\ref{i2}). 
Moreover, although we use a loopwise expansion to construct 
the fixed point (\ref{i4}) it can be regarded as ``non-Gaussian''
in the following sense: There exists a function $h^{\rm ren}(\,\cdot\,)$ 
for which $h^{\rm ren}_B(\,\cdot\,) = h^{\rm ren}(\,\cdot\,)$, so that 
one almost recovers conventional renormalizability. (A similar 
concept was recently employed in the context of T-duality \cite{BonCast}) 
However beyond one loop $h^{\rm ren}(\rho)$ {\it differs} from  
$h^{\rm beta}(\rho)$, in particular $\beta_h(h^{\rm ren}) \neq 0$. 
Thus if one was  
to explore the vicinity of this conventionally renormalizable
Lagrangian the fixed point (\ref{i4}) could not be seen and,
in view of Eq.~(\ref{zerotrace}) below, there was little chance to
impose the constraints. Finally we cannot resist mentioning 
the resemblance of (\ref{i4}) to the ``least coupling'' form of 
the dilaton functional proposed in \cite{PolyDam94}.

From a Wilsonian viewpoint the linearized flow 
around a fixed point encodes information about the critical 
manifold and the rate of approach to it. For the functional 
flow $\mu \,\ra \,\hbbar(\,\cdot\,,\mu)$ induced by (\ref{i3}) 
we write $\hbbar(\,\cdot\,,\mu) = h^{\rm beta}(\,\cdot\,) + 
\frac{\lb}{2\pi} \overline{\delta h}(\,\cdot\,,\mu) + O(\lb^2)$ and 
linearize in $\overline{\delta h}$. The resulting integro-differential 
equation can be solved analytically and reveals the following pattern 
\be
\overline{\delta h}(\rho,\mu) \;\rra \;0 \quad \mbox{for} \quad 
\left\{ \begin{array}{ll} 
\eps =+1 \;\;\,& \mbox{and}\;\;\,\mu \,\ra\, \infty\, \\
\eps =-1 \;\;\,& \mbox{and}\;\;\,\mu \,\ra\,  0\,.
\end{array} \right.
\label{i5}
\ee
The behavior (\ref{i5}) suggests ultraviolet stability of  
the fixed point for $\eps =+1$ and infrared stability for $\eps=-1$. 
Heuristically one can argue \cite{PTpaper} that the $\eps =+1$
case (where both Killing vectors are spacelike) captures the 
intuition behind a truncated version of the 4D functional 
integral for (Lorentzian or Euclidean) quantum gravity. 
The same is not true for the sector with one 
timelike and one spacelike Killing vector ($\eps =-1$). The existence 
of an ultraviolet fixed point in the non-renormalizable $\eps =+1$ 
theory therefore is in the spirit of the asymptotic safety 
scenario. To quote from \cite{Weinberg}: ``A theory is said to be 
asymptotically safe if the `essential' coupling parameters approach 
a fixed point as the momentum scale of their renormalization point 
goes to infinity.'' However there are also important differences:
Due to the complicated form of $\beta_h(h)$ no preferred parameterization
of $\hbbar(\,\cdot\,,\mu)$ in terms of numerical running couplings 
exist; the solutions $\overline{\delta h}$ can be grouped into classes with 
qualitatively different rate of decay (it is not always powerlike). 
In particular there is no obvious way to define the dimension 
of the critical manifold. 

In order for $h^{\rm beta}$ to qualify as a good fixed point 
one should also have a chance to impose the constraints when 
$h = h^{\rm beta}$. From the result (iii) described after Eq.~(\ref{cward}) 
we know that $T_{00} = \cH_0$ and $T_{01} = \cH_1$ can be defined 
as composite operators. Clearly a necessary condition for the quantum 
constraints to have a sufficiently large `kernel' on the 
state space generated by $\rho,\dd_{\mu} \sigma, J_{\mu}^i$
is that the trace anomaly vanishes. This is to say there
should exist an improvement potential $\Phi = f(\rho) + f_0 \sigma$ 
such that when $\nl T^{\mu}_{\;\;\mu} \nr$ is computed as a 
composite operator in the above scheme it vanishes modulo 
contributions proportional to the equations of motion operator.   
The counter terms relating $f_B(\rho_B)$ to $f(\rho)$ and 
those relating $\rho_B$ to $\rho$ depend on $h$. This triggers a 
non-autonomous inhomogeneous flow equation for a running 
$\fbar(\,\cdot\,,\mu)$. One might expect that when 
$\hbbar$ becomes stationary also $\fbar$ becomes stationary and 
defines the proper improvement potential at the fixed point. 
This is indeed the case, moreover 
\be
\mu \frac{d}{d \mu} \hbbar = 0 = \mu \frac{d}{d \mu} \fbar 
\quad \Longleftrightarrow \quad \nl T^{\mu}_{\; \mu} \nr =0 \,,
\label{zerotrace}
\ee
to all loop orders. 
That is, the trace anomaly of the improved energy momentum tensor 
vanishes precisely at the fixed point of the functional flow. 
The proof links the stationarity of $\fbar$ to a variant of the 
Curci-Paffuti relation \cite{CurciPaff,PTpaper}. 
The combinations $\nl \cH_0 \pm \cH_1 \nr$ are then expected 
to generate a 2D conformal algebra with formal central charge 
$c\!=\!4$. The value $c\!=\!4$ in itself has little significance because
the (sub-) state space generated by  $\rho,\dd_{\mu} \sigma$
has indefinite metric. The latter feature is not an artifact, it 
is directly related to the notorious ``conformal factor problem'' 
of 4D quantum Einstein gravity. As stressed in \cite{CJZ96}, even 
for free field doublets of opposite signature 
inequivalent quantizations exist which affect the value of $c$ 
through the choice of vacuum. For the 2-Killing reduction 
a complete construction of the physical state space is presumably
only possible in a bootstrap formulation as in \cite{qernst}.
Showing the equivalence of the present Lagrangian-based quantum
theory and the bootstrap theory is therefore a major desideratum.
For this already the construction of the first nonlocal quantum 
observable at the fixed point (following L\"{u}scher's strategy in the 
${\rm O}(3)$ sigma-model \cite{Lusch}) would be indicative. 
The Lagrangian construction on the other hand provides a direct 
link to 4D quantum Einstein gravity; in particular the fixed point 
in the truncated theory established here appears to be a necessary 
prerequisite for a fixed point in full quantum Einstein gravity. 
  
In this letter we investigated the 2-Killing reduction of pure 
Einstein gravity. Classically a large class of matter couplings can 
be included \cite{BMG} ranging from Einstein-Maxwell and dilaton-axion 
gravity \cite{Bakas} to $N\!=\!16$ supergravity \cite{Nic96}. The 
fixed point structure of all these systems can be studied 
along similar lines.\\
\mbox{}

\noindent 
{\tt Acknowledgments:} We wish to thank M.~Reuter, 
N.~Mohammedi, and D.~Maison for useful discussions.


\end{document}